\RequirePackage{fix-cm}
\documentclass{svjour3}                     
\smartqed  
\usepackage{graphicx}
\usepackage{array}
\usepackage{amsmath}
\usepackage{url}
\usepackage{amsfonts}
\usepackage{multirow}
\usepackage{tabularx}
\usepackage[misc]{ifsym}
\usepackage[caption=false]{subfig}
\journalname{World Wide Web}

\begin{document}

\title{Attribute-aware Explainable Complementary Clothing Recommendation
}


\author{Yang Li         \and
        Tong Chen       \and
        Zi Huang
}


\institute{Yang Li\at
                The University of Queensland, Brisbane, Australia \\
                \email{yang.li@uq.edu.au}\\
           \and
          Tong Chen \at
                The University of Queensland, Brisbane, Australia \\
                \email{tong.chen@uq.edu.au}\\
            \and
          \Letter \ Zi Huang \at
                The University of Queensland, Brisbane, Australia \\
                \email{huang@itee.uq.edu.au}\\
}

\date{Received: date / Accepted: date}

\maketitle

\begin{abstract}
Modelling mix-and-match relationships among fashion items has become increasingly demanding yet challenging for modern E-commerce recommender systems. When performing clothes matching, most existing approaches leverage the latent visual features extracted from fashion item images for compatibility modelling, which lacks explainability of generated matching results and can hardly convince users of the recommendations.
Though recent methods start to incorporate pre-defined attribute information (e.g., colour, style, length, etc.) for learning item representations and improving the model interpretability, their utilisation of attribute information is still mainly reserved for enhancing the learned item representations and generating explanations via post-processing. As a result, this creates a severe bottleneck when we are trying to advance the recommendation accuracy and generating fine-grained explanations since the explicit attributes have only loose connections to the actual recommendation process. This work aims to tackle the explainability challenge in fashion recommendation tasks by proposing a novel Attribute-aware Fashion Recommender (AFRec). Specifically, AFRec recommender assesses the outfit compatibility by explicitly leveraging the extracted attribute-level representations from each item's visual feature. The attributes serve as the bridge between two fashion items, where we quantify the affinity of a pair of items through the learned compatibility between their attributes. Extensive experiments have demonstrated that, by making full use of the explicit attributes in the recommendation process, AFRec is able to achieve state-of-the-art recommendation accuracy and generate intuitive explanations at the same time.
\end{abstract}

\begin{figure}[t]
\centering
\includegraphics[width=0.7\textwidth]{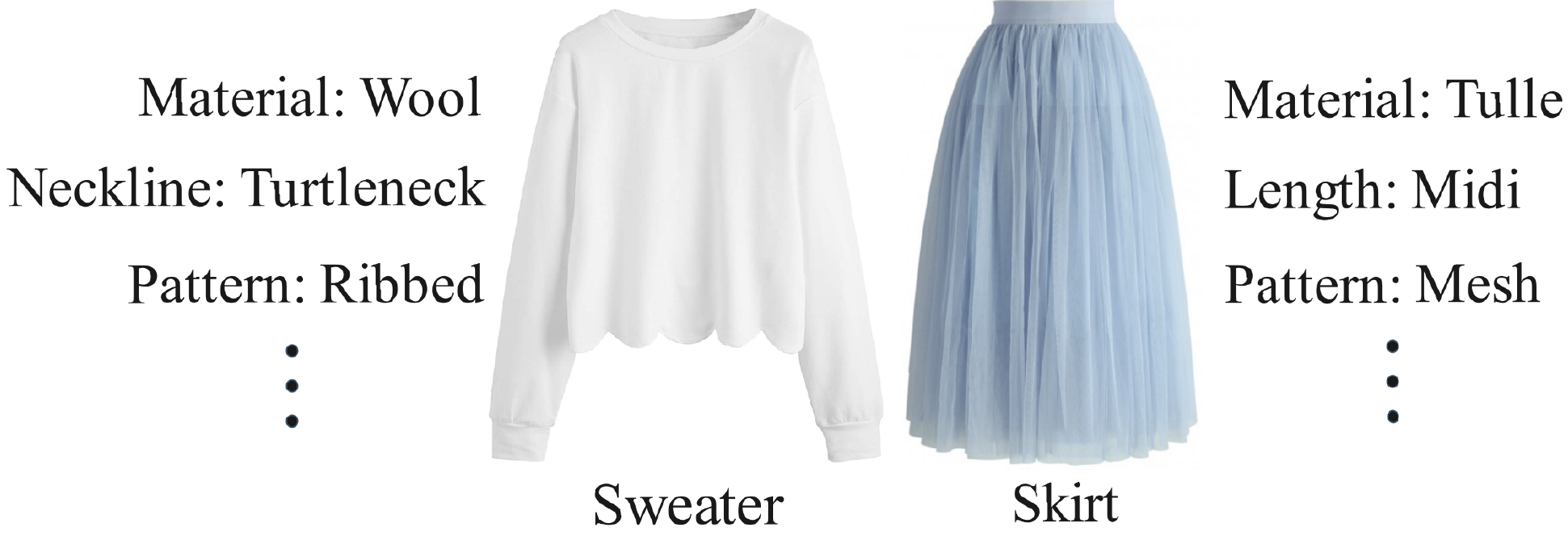}
\caption{An example of clothing attributes}
\label{fig:attribute_example}
\end{figure}

\section{Introduction}
\label{intro}
The advancement of modernisation attracts rapidly growing attention to fashion. A wide range of fashion-focused social websites have emerged in recent decades, such as Polyvore$\footnote{https://www.polyvore.com}$ and ShopLook$\footnote{https://www.shoplook.io}$. With an overwhelming amount of product choices, customers nowadays are craving for personal advice on outfit matching and recommendation of the most suitable item for their wardrobes, which brings in a great opportunity of designing automated tools for measuring fashion compatibility. 

The recent research in fashion domain evolves from fundamental clothing recognition \cite{LiuLQWT16,ZhangZYW20}, style understanding \cite{Chiu19} to aesthetic and compatibility analysis \cite{TangsengYO18,LiCZL17,HanWJD17,song2019compatibility,LiLH20}. Learning compatibility relationships is a challenging and sophisticated task, as whether two clothes (e.g., top and bottom clothes) are a good match is usually determined by a complex mixture of various factors. A large body of work on this task models compatibility notions by computing latent representations for a given pair of items, then modelling the similarity between items via those representations \cite{TangsengYO18,LiCZL17,HanWJD17,li2020fashion}. In this regard, latent factor models, especially deep models \cite{HanWJD17,SongFLLNM17} have commonly demonstrated promising recommendation accuracy. However, the main drawback of these latent factor methods is that the recommendation process is non-transparent to users, making it hard for users to justify the reasons behind successfully matched clothes. In the real-world scenario, users usually not only want to know whether two outfits are compatible or not but also would like to understand the major factors that lead to the failure or success of matching. 

Though visual explanations (usually made with attention) are offered in some recent methods to reveal a model's inner mechanism and perform model validation \cite{kang2019complete}, however, they are less helpful for convincing users of the generated clothes matching results and making detailed explanations beyond only the appearance of items. In fact, as illustrated in Figure \ref{fig:attribute_example}, the property of a fashion item can be further decomposed into multiple fine-grained attributes (e.g., shape, colour, pattern, material, etc.), which are highly relevant when users are shopping for clothes. To enhance the model interpretability, some work attempts to incorporate information of pre-defined attributes of clothes when modelling clothes compatibility. However, despite the availability of attribute information, the attributes are only involved in the recommendation process in the form of latent features of items, thus giving up the rich compatibility signals between explicit attributes and making the generated explanation coarse-grained. For example, \cite{HanSYWN19} generates explanations by post-processing the associated attributes after a recommendation is made, making the attribute-wise explanations loosely connected to the actual recommendation results. Meanwhile, \cite{Yang0WMF0C19} requires pretraining an individual decision tree before meaningful attribute combinations can be used for clothes matching and interpretation, and the quality of both recommendation and explanation is highly dependent on the selected decision tree model. 

To alleviate the aforementioned limitations of previous work, we introduce our Attribute-aware Fashion Recommender (AFRec), which makes full use of explicit attribute information to mimic a human's decision-making process where the compatibility of two clothes are usually determined by comparing various attributes of both items. Specifically, taking the images of a pair of clothes as the input, AFRec utilises a pretrained convolutional neural network (CNN) to extract visual features from both clothes. Then, we design an innovative semantic attribute extractor that automatically maps each item to a group of attribute representations. Unlike existing attribute-based methods that directly fuse extracted attributes into a unified representation for each item \cite{Yang0WMF0C19,HanSYWN19}, we disentangle the straightforward item-item affinity into the explicit attribute-attribute compatibility. To achieve this, we propose a novel attribute-wise reciprocal attention module, where the affinity between two items is conditioned on the inherent compatibility of each attribute pair as well as each item's performance across all attributes. This enables AFRec to precisely bridge two complementary clothes with fine-grained attributes. Moreover, the pairwise attribute compatibility scores allow AFRec to provide intuitive attribute-level explanations on the recommendation results. 

Our main contributions are summarised as follows:
\begin{itemize}
	\item We approach an emerging and important research problem - explainable complementary clothes recommendation from a different view, i.e., using attribute-level compatibility to bridge two complementary clothes.
	
	\item We propose Attribute-aware Fashion Recommender (AFRec), a novel model that explores the fine-grained attribute-level collocation via a CNN-based semantic attribute extractor, which is followed by an innovative attribute-wise attentive compatibility modelling paradigm for clothes matching. 
	
	\item We extensively evaluate AFRec on two benchmark datasets, where the results suggest that it is able to outperform state-of-the-art baselines and generate intuitive explanations at the same time.
\end{itemize}

\section{Related Work}
In recent years, a variety of recommender systems have been developed in various areas, such as POI recommendation \cite{yin2013lcars,wang2015geo,yin2016adapting,LiLZSC19,li2019context}, sequential recommendation \cite{chen2019air,chen2020sequence,yin2019social,guo2021gcn} and complimentary recommendation \cite{chen2020try}. However, the conventional recommender systems are mainly developed using item IDs and textual information, which fail to leverage the important visual signals for recommendation. The rapid development of computer vision area has significantly promoted various visual-based applications, such as image retrieval \cite{LuoZWCHX18,ZhangXLLH19,luo2020collaborative,wang2020deep,zhang2020inductive,ChenLLNZX20,zhang2021high,Zhang2021}, visual understanding \cite{wang2018look,luo2019curiosity,chen2020rethinking}, and visual domain adaptation \cite{luo2020progressive,luo2020adversarial,wang2020prototype}. This also has largely facilitated the studies in the fashion area. The existing work on recommending complementary clothing items mainly utilises the visual signals extracted from the product image data to model the visual correlations between items and user preferences. McAuley et al. \cite{McAuleyTSH15} propose to use Low-rank Mahalanobis Transformation to learn a latent style space for minimising the distance between matched clothing item embeddings and maximising that of mismatched ones. Veit et al. \cite{VeitKBMBB15} employ the Siamese CNNs to learn a metric for compatibility measurement in an end-to-end manner. Some researchers argue that the complex compatibility relationships cannot be captured by directly learning a single latent space. He et. al \cite{HePM16} propose to learn a mixture of multiple metrics with weight confidences to model the relationships between heterogeneous items. Veit et al. \cite{VeitBK17} propose Conditional Similarity Network, which learns disentangled item features whose dimensions can be used for separate similarity measurements. Li et al. \cite{LiCZL17} use an encoder to fuse features from multimodal inputs and adopt pooling techniques to get a single representation of an outfit for compatibility measurement. Vasileva et al. \cite{VasilevaPDRKF18} claim that respecting type information has important consequences. Thus, they build type-wise trainable mask embeddings and use them to attend on different latent aspects when measuring different kinds of top-bottom pairs. Similarly, Yang et al. \cite{YangMLWC19} introduce a translation-based type-aware model, which learns type-specific embeddings to connect compatible item embedding pairs. Different from the previous category-aware work \cite{VasilevaPDRKF18,YangMLWC19}, instead of learning either mask or categorical relation embeddings, we build category-specific weight matrices in AFRec, which help the model to focus on different latent aspects for attribute representation pairs in different categorical groups.

However, there are some voices arguing that these previous methods suffer from limited interpretability. Han et al. \cite{HanSYWN19} propose a Bayesian Personalised Ranking (BPR) framework named PAICM that adopts NMF to learn the latent attribute-level prototype embeddings for both compatible and incompatible outfits. Thus, the model could provide a recommendation explanation by comparing the item-level embedding with the closest prototype embedding. However, since the interpretability of this method highly relies on the quality of the learned prototype embeddings, the model is sensitive to the number of defined prototypes. Xun et al. \cite{Yang0WMF0C19} propose to draw harmonious matching rules through a deep decision tree for the explainability of the recommendation model. Another explainable fashion recommendation model \cite{LinRCRMR20} learns to generate review comments by an attentive RNN-based decoder using the fused item-level embeddings. Nevertheless, these approaches either require abundant well-annotated attribute labels of each item for matching rule mining or user-generated reviews for training the explanation generation component. This impedes the practicality of those methods on most fashion datasets, where only a short textual description is available for each clothing item. Different from those methods, our model innovatively captures the fine-grained pairwise interactions at the attribute level, which provides an explicit and clear explanation by automatically concentrating on the most important attribute factors in a given compatible/incompatible outfit pair.


\begin{figure}[t]
\centering
\includegraphics[width=\textwidth]{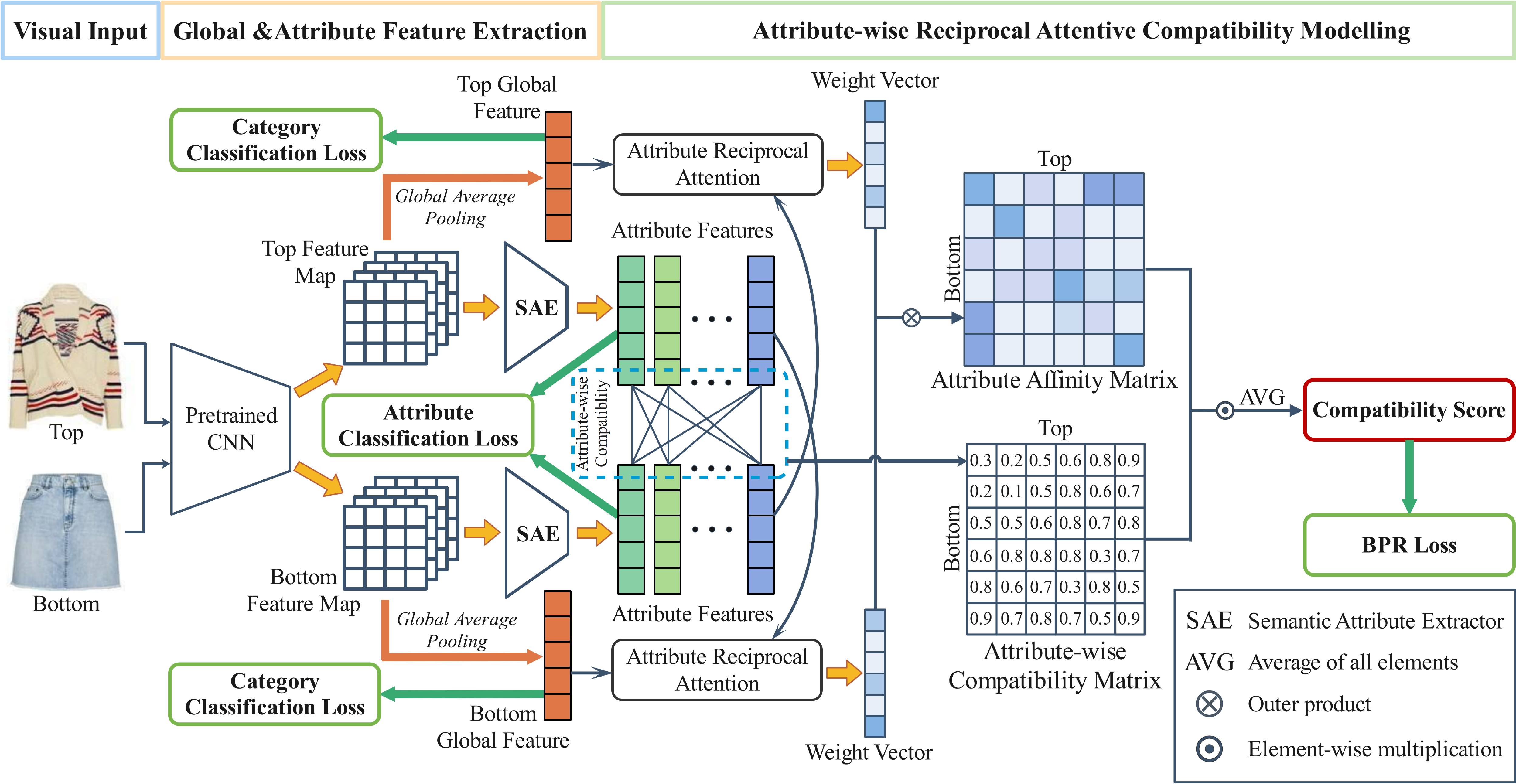}
\caption{An overview of our proposed AFRec model}
\label{fig:model_overview}
\end{figure}

\section{Problem Formulation}
In this paper, we focus on the widely studied problem of matching top and bottom clothes \cite{LiLH20,Yang0WMF0C19,chenaaai18,LiuFDXHHY14,LiLH20}, while our approach can be easily generalised to other types of clothes matching problems. Let us use $\mathcal{T}=\{t_1, t_2, ..., t_{N^t}\}$, $\mathcal{B}=\{b_1, b_2, ..., b_{N^b}\}$, $\mathcal{A}=\{a_1, a_2, ..., a_K\}$ and $\mathcal{C}=\{c_1,c_2,...,c_{|\mathcal{C}|}\}$ to denote the set of top images, bottom images, attributes and item categories in the dataset. $\mathcal{D}$, where $N^t$, $N^b$, $K$ and $L$ are the total numbers of tops, bottoms, attributes and item categories, respectively.
Bold lowercase letters and bold uppercase letters are used to indicate embedding vectors and matrices, respectively. 

In this work, we target at modelling outfit compatibility as well as exploring the explainability of the generated recommendations. Formally, given an arbitrary top-bottom pair $(t_i, b_j)$, our model is able to utilise the attribute information $\mathcal{A}$ associated with each item to distinguish whether $t_i$ and $b_j$ is a qualified match or not. In the case of the ranking task, our model is expected to generate the highest ranking score for a ground truth item pair than a non-matching item pair.

\section{Proposed Approach}
As discussed in Section \ref{intro}, most existing work models fashion compatibility by measuring the similarity between fashion items' latent representations, where the meaning of the features is incomprehensible to users. As a result, they could hardly provide convincing explanations for their predictions. To address this limitation, we propose an attribute-aware fashion recommender, namely AFRec, which supports comprehensive clothes matching and reasoning at the attribute level. The workflow of AFRec is shown in Figure \ref{fig:model_overview}. In this section, we first introduce the global and attribute-specific representation extraction procedure. Then, we describe our designed attribute reciprocal attention mechanism, which fully explores the complementary correlations between the top and bottom attributes for compatibility modelling. Finally, we give the learning objective for training our model.

\subsection{Item Visual Feature Extraction}\label{sec:itemvisfea}
As illustrated in the left part of Figure \ref{fig:model_overview}, we first utilise a pretrained CNN to extract high-level visual features from the raw input images. Considering both performance and computational complexity, we adopt ResNet-18 \cite{HeZRS16} pretrained on ImageNet dataset \cite{RussakovskyDSKS15} as the backbone module. Accordingly, for image $t_i$/$b_j$ that are of the size $224 \times 224$ with 3 colour channels, the feature maps output from the pretrained CNN can be represented as $\textbf{F}_{t_i} \in \mathbb{R}^{D\times7\times7}$ and $\textbf{F}_{b_j} \in \mathbb{R}^{D\times7\times7}$, where $D$ is the output dimension size ($D=512$ in a typical ResNet-18), and $7\times7$ denotes the output feature map size, i.e., height $\times$ width.

\textbf{Generating Global Item Embeddings.} To compress the visual feature maps into a compact item embedding, we use two sets $\mathcal{F}_{t_i}$ and $\mathcal{F}_{b_j}$ to collect all $7\times 7 = 49$ $D$-dimensional feature vectors, i.e.,  $\mathcal{V}_{t_i} = \{\textbf{v}^{t_i}_1,\textbf{v}^{t_i}_2,...,\textbf{v}^{t_i}_{49}\}$ and $\mathcal{V}_{b_j} = \{\textbf{v}^{b_j}_1,\textbf{v}^{b_j}_2,...,\textbf{v}^{b_j}_{49}\}$ where $\textbf{v} \in \mathbb{R}^D$ corresponds to one feature in the feature map. Then, the global embedding vectors of items $t_i$ and $b_j$
can be obtained by feeding $\mathcal{V}_{t_i}$ and $\mathcal{V}_{b_j}$ into a global average pooling layer:
\begin{equation}
\label{eq:globalitememb}
\textbf{v}_{t_i} = \frac{1}{49}\sum_{n=1}^{49}{\textbf{v}^{t_i}_n}, \,\,\,\,\,\,
\textbf{v}_{b_j} = \frac{1}{49}\sum_{n=1}^{49}{\textbf{v}^{b_j}_n},
\end{equation}
where $\textbf{v}_{t_i}^{global}$, $\textbf{v}_{b_j}^{global} \in \mathbb{R}^{D}$ denote the global feature embedding for $t_i$ and $b_j$, respectively.

\textbf{Fine-tuning Pretrained CNN.} As the pretrained ResNet-18 is not originally designed for attribute-aware fashion recommendation, we fine-tune this CNN module with an item categorisation task. The rationale is that, fashion items of different categories tend to demonstrate different distributions over attributes. For instance, ``sleeve length'' is an important attribute for shirts and sweaters, while people tend to pay more attention to the ``waistline'' of a dress. This requires the model to focus on different attributes when handling different types of clothes. Therefore, to generate category-sensitive and more discriminative item embeddings to better guide the subsequent attribute extraction procedure, we design an additional item classification task with cross-entropy loss, which is used to fine-tune the pretrained CNN module: 
\begin{equation}
    \begin{aligned}
        \widehat{\textbf{y}}_{item} &= \textrm{softmax}(\textbf{W}^{cat}\textbf{v}^{global}_{item} + \textbf{b}^{cat}),\\
        \mathcal{L}_{category} &= - \sum_{\forall item \in \mathcal{T}\cup\mathcal{B}} \textbf{y}^{\top}_{item} \log (\widehat{\textbf{y}}_{item}),
    \end{aligned}
    \label{eq:category_loss}
\end{equation}
where $\textbf{W}^{cat} \in \mathbb{R}^{|\mathcal{C}|\times D}$ and $\textbf{b}^{cat} \in \mathbb{R}^{|\mathcal{C}|}$ are the weight and bias of the classifier, $\widehat{\textbf{y}}_{item} \in \mathbb{R}^{|\mathcal{C}|}$ is the predicted probability distribution over all item categories, and $\textbf{y}_{item}$ is an one-hot encoding of each item's ground truth category label.

\subsection{Semantic Attribute Representation Extraction}
\begin{figure}[t]
\centering
\includegraphics[width=0.8\textwidth]{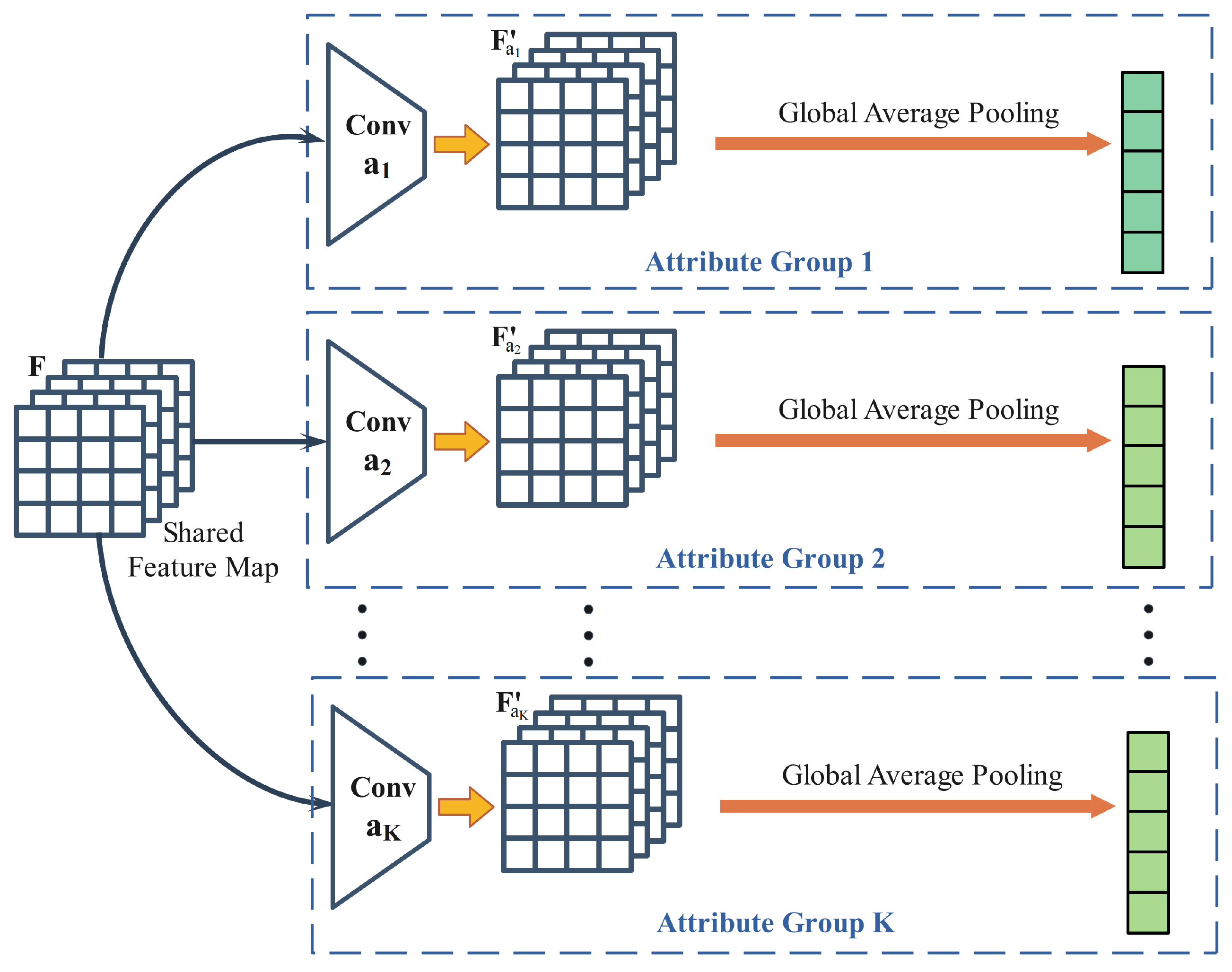}
\caption{An overview of Semantic Attribute Extractor (SAE)}
\label{fig:sae_overview}
\end{figure}
On e-commerce websites, on top of visual information (i.e., images), a fashion garment usually has a textual description at the same time. This allows us to effectively summarise meaningful item attributes such as shape, pattern and style. With a pre-defined item attribute set $\mathcal{A}$, we propose a CNN-based semantic attribute extractor (SAE) for meaningful attribute-specific region localisation and representation generation in a weakly supervised manner. Previous attribute-aware solutions \cite{Yang0WMF0C19,HanSYWN19} learns universal representation for every single attribute, and use the combinatorial feature of different attributes for item representation learning. However, using fixed attribute representations lacks adequate flexibility as each item may exhibit different characteristics towards each attribute. Hence, in AFRec, we allow each item to have its unique representation regarding an attribute $a_k$, which is learned in an attribute-specific feature space. 

As illustrated in Figure \ref{fig:sae_overview}, the extracted feature map $\textbf{F} \in \mathbb{R}^{D\times 7 \times 7}$ is shared over all attribute-specific blocks (each block is marked by blue lines in Figure \ref{fig:sae_overview}). There are $K$ blocks defined in SAE corresponding to $K$ fashion attributes. For the $k$-th attribute $a_k \in \mathcal{A}$, we adopt an independent convolutional layer whose kernel size is of $D \times 1 \times 1$ to transform the visual feature map $\textbf{F}$ to $\textbf{F}'_k \in \mathbb{R}^{D\times 7 \times 7}$. Note that the convolutional layer in each attribute block has a unique set of parameters. Then, with a global average pooling operation as in Eq.(\ref{eq:globalitememb}), we can obtain attribute $a_k$'s embedding vector $\textbf{a}_k \in \mathbb{R}^{D}$. Similarly, the same attribute extraction scheme is applied in all other blocks. Accordingly, for both items $t_i$ and $b_j$, we stack all $K$ attribute representations obtained from SAE into two $K\times D$ matrices, i.e., $\textbf{A}_{t_i} = [\textbf{a}_1^{t_i}, \textbf{a}_2^{t_i}, ..., \textbf{a}_K^{t_i}]$ and  $\textbf{A}_{b_j} = [\textbf{a}_1^{b_j}, \textbf{a}_2^{b_j}, ..., \textbf{a}_K^{b_j}]$. Intuitively, $\textbf{A}_{t_i}$ and $\textbf{A}_{b_j}$ can be viewed as two attribute-aware feature matrices representing $t_i$ and $b_j$. 

Apparently, we can directly optimise each $\textbf{a}_k$ within $\textbf{A}_{t_i}$ and $\textbf{A}_{b_j}$ using downstream clothes matching tasks. However, to ensure sufficient expressiveness of the learned attribute representation $\textbf{a}_k$, we further introduce a prediction task in the SAE layer. To be specific, for each attribute, e.g., $a_k = $``colour'', we can obtain the ground truth label (i.e., value) from the corresponding item, e.g., ``colour''$\rightarrow$``white''. Suppose for the $k$-th attribute, there are $N^k$ possible values, then we can use one-hot encoding $\textbf{z}_{k}^{item}$ to label the observed value on $item \in \mathcal{T}\cup \mathcal{B}$. In a similar vein to Section \ref{sec:itemvisfea}, we perform attribute value prediction with cross-entropy loss:
\begin{equation}
    \begin{aligned}
        \widehat{\textbf{z}}^{item}_k &= \textrm{softmax}(\textbf{W}^{attr}_k\textbf{a}_k^{item} + \textbf{b}^{attr}_k),\\
        \mathcal{L}_{attribute} &= - \sum_{\forall item \in \mathcal{T}\cup\mathcal{B}}\,\sum_{k=1}^{K} \textbf{z}^{item\top}_k \log (\widehat{\textbf{z}}^{item}_k),
    \end{aligned}
    \label{eq:attr_loss}
\end{equation}
where $\textbf{W}^{attr}_k \in \mathbb{R}^{K\times D}$ and $b^{attr}_k \in \mathbb{R}^{N^k}$ are the weight and bias of the classifier for the $k$-th attribute, $\widehat{\textbf{z}}^{item}_k \in \mathbb{R}^{N^k}$ denotes the item's estimated probability distribution over all possible values of attribute $a_k$. By optimising $\mathcal{L}_{attribute}$, we can effectively enforce the attribute embedding learned in each block to be a high-quality reflection on the $k$-th attribute of the target item.

\subsection{Attribute-wise Reciprocal Attention}
As a common practice, people tend to consider the different combinations of top and bottom attributes when choosing clothes to wear. For example, when a person wants to find a pair of pants to match his/her T-shirt, he/she may think about whether the colour and the pattern of the pants are compatible with the T-shirt. Inspired by the recent advances of attention mechanism in computer vision \cite{HanGZZ18,ZhouKLOT16,JetleyLLT18}, we propose an attribute-wise reciprocal attention mechanism for clothes matching. In particular, for top $t_i$'s attribute representation $\textbf{a}^{t_i}_k \in \textbf{A}_{t_i}$, an attention score $s^{t_i}_k$ representing its importance to the bottom $b_j$ can be computed via the following:
\begin{equation}
    s^{t_i}_k = \textbf{w}^{\top}\textrm{tanh}(\textbf{W}_1 \textbf{a}^{t_i}_k + \textbf{W}_2 \textbf{v}_{b_j}^{global}),
\end{equation}
where $\textbf{w} \in \mathbb{R}^D$ carries the projection weight, while $\textbf{W}_1 \in \mathbb{R}^{D \times D}$ and $\textbf{W}_2 \in \mathbb{R}^{D \times D}$ are two weight matrices. Then, a normalised reciprocal attention weight $\alpha^{t_i}_k$ for $t_i$'s attribute $a^{t_i}_k$ is calculated as follows:
\begin{equation}
    \alpha^{t_i}_k = \frac{\exp(s^{t_i}_k)}{\sum^{K}_{k=1} \exp(s^{t_i}_k)}.
\end{equation}
Now, we can get $t_i$'s reciprocal attribute attention weight vector $\textbf{v}_{t_i}^{attn}=[\alpha^{t_i}_1,\alpha^{t_i}_2,$\\$...,\alpha^{t_i}_{K}] \in \mathbb{R}^{K}$, where the value of $k$-th dimension in $\textbf{v}_{t_i}^{attn}$ represents $t_i$'s performance on the $k$-th attribute regarding the bottom $b_j$. Similarly, we can also obtain $b_j$'s attribute attention vector $\textbf{v}_{b_j}^{attn}$:
\begin{equation}
  \begin{aligned}
    s^{b_j}_k &= \textbf{v}_{attn}^{\top}\textrm{tanh}(\textbf{W}_1 \textbf{a}^{b_j}_k + \textbf{W}_2 \textbf{v}_{t_i}^{global}),\\        
    \alpha^{b_j}_k &= \frac{\exp(s^{b_j}_k)}{\sum^{K}_{k=1} \exp(s^{b_j}_k)},\\
    \textbf{v}_{b_j}^{attn}&=[\alpha^{b_j}_1, \alpha^{b_j}_2,...,\alpha^{b_j}_{K}] \in \mathbb{R}^{K}.
  \end{aligned}
\end{equation}

So far, $\textbf{v}^{attn}_{t_i}$ and $\textbf{v}^{attn}_{b_j}$ can be viewed as attribute-aware representations of $t_i$ and $b_j$, which are respectively conditioned on each other. 

\subsection{Explicit Attribute-aware Compatibility Modelling}
With the obtained attribute representations $\mathbf{A}_{t_i}$ and $\mathbf{A}_{b_j}$, we then perform the compatibility prediction in a pairwise manner, which is illustrated in the right part of Figure \ref{fig:model_overview}. To be specific, for each top-bottom pair $(t_i, b_j)$, we first project their attribute representations into a category-specific space via linear transformation, then we obtain a compatibility matrix $\mathbf{M}^{compat}_{ij} \in \mathbb{R}^{K\times K}$ by calculating an affinity score between every pair of attribute-wise representations $(\mathbf{a}^{t_i}_k, \mathbf{a}^{b_j}_{k'})$. To allow for efficient computation, the matrix-level computation is written as:
\begin{equation}
    \label{eq:compat_calculation}
    \mathbf{M}^{compat}_{ij} = (\mathbf{A}_{t_i}\mathbf{W}^{c_ic_j}) \mathbf{W}^{compat} (\mathbf{A}_{b_j}\mathbf{W}^{c_ic_j})^{\top} \in \mathbb{R}^{K \times K},
\end{equation}
where $\mathbf{W}^{c_ic_j} \in \mathbb{R}^{D \times D}$ is a category-specific weight matrix, here, $c_ic_j$ denotes the pair of $(t_i, b_j)$'s categorical labels, $\mathbf{W}^{compat} \in \mathbb{R}^{D\times D}$ is a transformation matrix that aligns the feature spaces for both attribute-wise feature matrices for compatibility measurement. Each element $m^{compat}_{kk'} \in \mathbf{M}_{ij}^{compat}$ results from the dot product between the intrinsic attribute representations $\mathbf{a}^{t_i}_k$ and $\mathbf{a}^{b_j}_{k'}$. Hence, $m^{compat}_{kk'}$ can be viewed as the inherent compatibility score for attribute pair $(a_k, a_{k'}$, which is learned with the contexts given by the $(t_i, b_j)$ pair. A higher score means that two attributes are closely correlated for clothes matching, e.g., attributes ``bottom length'' and ``waistline'' are usually bounded when matching the sizing of clothes. 

Moreover, recall that we have also obtained the attention vectors $\mathbf{v}^{t_i}_{attn}, \textbf{v}^{b_j}_{attn} \in \mathbb{R}^K$ for both items. Intuitively, the $k$-th element in $\textbf{v}^{t_i}_{attn}$/$\textbf{v}^{b_j}_{attn}$ indicates the performance of item $t_i$/$b_j$ on a specific attribute $a_k$. By  performing an outer product between those two vectors, we can enumerate over all the pairwise combinatorial effect between $t_i$ and $b_j$'s attribute-wise performance:
\begin{equation}
    \textbf{M}_{ij}^{affinity} = \textbf{v}_{t_i}^{attn} \otimes \textbf{v}_{b_j}^{attn} \in \mathbb{R}^{K \times K},
\end{equation}
where $\otimes$ is an outer product operator, and  $\textbf{M}_{ij}^{affinity} \in \mathbb{R}^{K\times K}$ is an explicit affinity matrix where a large element $m^{affinity}_{kk'} \in \textbf{M}_{ij}^{affinity}$ indicates that $t_i$ and $b_j$ are respectively performing well on attributes $a_k$ and $a_k'$, yielding a higher affinity score between their explicit attributes.

Then, a weighted compatibility score matrix $\textbf{M}^{weighted\_compat}_{ij}$ for $t_i$ and $b_j$ can be obtained via an element-wise multiplication:
\begin{equation}
    \textbf{M}^{weighted\_compat}_{ij} = \textbf{M}_{ij}^{affinity} \odot \textbf{M}^{compat}_{ij} \in \mathbb{R}^{K \times K},
\end{equation}
where $\odot$ is an element-wise multiplication operator. Through element-wise multiplication, it is clear that a large score $m_{kk'} \in \textbf{M}^{weighted\_compat}_{ij}$ can be obtained only if $m^{compact}_{kk'}$ and $m^{affinity}_{kk'}$ are both high. Hence, for $t_i$ and $b_j$, a large $m_{kk'}$ means: (1) their attributes $a_k$ and $a_k'$ complement each other; and (2) $t_i$ and $b_j$ have ideal performance on $a_k$ and $a_k'$, respectively. In this way, AFRec is able to give an explicit explanation on which pairs of attributes are most critical and have the most positive or negative impacts on the outfit. Furthermore, by incorporating fine-grained attribute-level affinity, we have higher chances of improving the recommendation accuracy, because the complementary information from different attribute views offers richer signals for identifying compatible clothes.

Eventually, the final compatibility score between $t_i$ and $b_j$ $\widehat{y}^{compat}_{ij}$ is a scalar derived by summing up all elements in $\textbf{M}^{weighted\_compat}_{ij}$:
\begin{equation}
    \widehat{y}^{compat}_{ij} = \sum_{k=1}^{K}\sum_{k'=1}^{K} m_{kk'},\,\,\,\, 
    m_{kk'} \in \textbf{M}^{weighted\_compat}_{ij}.
    \label{eq:weight_compat_matrix}
\end{equation}

\subsection{Model Optimisation}
In a sense, only the positive top-bottom outfit pairs created by fashion experts are available in the dataset, while the negative pairs are unknown. Thus, we employ a soft-ranking loss function, namely Bayesian Personalised Ranking (BPR) \cite{RendleFGS09} to explore the implicit relations between tops and bottoms. Specifically, for each observed top-bottom pair $(t_i, b_j)$, we can generate two corrupted pairs $(t_i, b_{j'})$ and $(t_{i'}, b_j)$ by replacing either the top or bottom with an unobserved one. Then, based on the assumption that the observed pairs should be ranked higher than unobserved ones, we have:
\begin{equation}
	\mathcal{L}_{bpr} = -\sum_{(i,j,j') \in \mathcal{D}}\ln\Big{(}\sigma(\widehat{y}_{ij}^{compat} - \widehat{y}_{ij'}^{compat} )\Big{)},
\end{equation}
where $\mathcal{D}$ denotes all the training instances and $\sigma$ is a sigmoid function. Note that we have omitted the corrupted instance for tops (i.e., $(i', i, j)$) to be succinct. 

Ultimately, the final objective function of AFRec is formulated as follows:
\begin{equation}
	\mathcal{L} = \mathcal{L}_{bpr} + \mathcal{L}_{category}  + \mathcal{L}_{attribute}.
	\label{eq:final_loss}
\end{equation}

\section{Experiments}
To verify the effectiveness of our proposed model, we conduct extensive experiments on two real-world benchmark datasets, i.e., FashionVC and PolyvoreMaryland. We first describe the details of experimental settings and then give a comprehensive analysis according to the experimental results.

\subsection{Experimental Settings}
\textbf{Datasets.} We conduct experiments on two public fashion benchmark datasets, namely FashionVC and PolyvoreMaryland. \textbf{FashionVC}\footnote{\url{https://drive.google.com/file/d/1lO7M-jSWb25yucaW2Jj-9j_c9NqquSVF/view}} is released by Song et al. \cite{SongFLLNM17}, which consists of 20,726 outfits including 14,871 top item images and 13,663 bottom item images, created by fashion domain experts. Each clothing item in the dataset corresponds to an image, a text title and a category label. \textbf{PolyvoreMaryland}\footnote{\url{https://drive.google.com/drive/folders/0B4Eo9mft9jwoVDNEWlhEbUNUSE0}} is created by Han et al. \cite{HanWJD17}, which has which has 21,889 outfits in total.
We use images for visual information extraction, and characterise the attributes based on each item's title and category label. All the attributes and examples of their corresponding values are summarised in Table \ref{tab:attribute_labels}. We randomly split the data by 80\%:10\%:10\% for training, validation and test, respectively.

\begin{table}
\caption{A summarisation of all attributes and their corresponding values. We also report the attribute classification accuracy of AFRec on these attributes}
\label{tab:attribute_labels}
\setlength\tabcolsep{3.2pt}
\renewcommand{\arraystretch}{2}
\begin{tabularx}{\textwidth}{llll}
\cline{1-4}
\multirow{2}{*}{Attribute} & \multirow{2}{*}{Attribute Values} & \multicolumn{2}{c}{Classification Accuracy (\%)}\\
\cline{3-4}& & FashionVC & Polyvore \\ \cline{1-4}
 
Category & T-shirt, sweatshirts, cardigans sweaters, ... & 98.2 & 87.3 \\ \cline{1-4}
Texture & cotton, fur, leather, velvet, metallic, ...  &  99.4 & 80.3\\ \cline{1-4}
Style & patchwork, woven, slit, cuffed, sheer, raw, ... & 99.4 & 84.6\\ \cline{1-4}
Pattern & chino, houndstooth, striped, grid, crochet, ... & 99.4 & 84.0\\ \cline{1-4}
Neckline & scoop neck, v-neck, high-neck, tie-neck, ... & 99.5 & 96.2\\ \cline{1-4}
Sleeve Type & sleeveless, long sleeve, short sleeve, ... & 98.8 & 92.3\\ \cline{1-4}
Shape & oversized, stretch, skinny, loose, ... & 97.7 & 73.1\\ \cline{1-4}
Waistline & high waist, mid waist, low waist, ... & 99.8 & 92.0\\ \cline{1-4}
Bottom Leg & harem, straight-leg, cropped & 99.9 & 95.2\\ \cline{1-4}
Bottom Length & maxi, mini, midi & 99.6 & 96.1\\ \cline{1-4}

\end{tabularx}
\end{table}

\subsection{Baseline Methods}
 We compare our model AFRec with several state-of-the-art models for complementary clothing recommendation.
 \begin{itemize}
 	\item \textbf{SiameseNet} \cite{VeitKBMBB15}: The approach models compatibility by minimising the Euclidean distance between clothes pairs and making the incompatible ones far apart within a unified compatibility latent space through a contrastive objective function.
 	\item \textbf{Monomer} \cite{HePM16}: The approach models fashion compatibility with a mixture of distances computed from multiple latent spaces.
 	\item \textbf{BPR-DAE} \cite{SongFLLNM17}: The approach models the overall matching knowledge through an inner product of the top and bottom visual representations.
 	\item \textbf{TripletNet} \cite{chenaaai18}: 
 	This is the state-of-the-art approach that captures the complementary relations among different fashion items with a triplet objective function.
 	\item \textbf{TA-CSN} \cite{VasilevaPDRKF18}: This is a category-aware method that considers item categorical awareness by generating category-specific masks added upon item visual embeddings, which helps the model to concentrate on different latent aspects when modelling items from different categories.
 	\item \textbf{PAICM} \cite{HanSYWN19}: It is the state-of-the-art attribute-aware approach that employs non-negative matrix factorisation to explore the pairwise compatibility at the attribute level.
 \end{itemize}
 
  \begin{table}[t]
	\caption{Performance comparison between our proposed AFRec and other baseline methods}
	\setlength\tabcolsep{3.3pt}
	\renewcommand{\arraystretch}{2}
	\begin{tabularx}{\textwidth}{llllllp{1pt}llllr}
		\cline{1-12}
		\multirow{3}{*}{Methods} & \multicolumn{5}{l}{FashionVC} & & \multicolumn{5}{l}{PolyvoreMaryland} \\
		\cline{2-6} \cline{8-12}
		 & \multirow{2}{*}{AUC} & \multicolumn{4}{l}{HR$@$K} & & \multirow{2}{*}{AUC} & \multicolumn{4}{l}{HR$@$K} \\
		\cline{3-6} \cline{9-12} &
		  & K=5 & K=10 & K=20 & K=40 & & & K=5 & K=10 & K=20 & K=40\\
		\cline{1-12}
		SiameseNet    & 0.604     & 0.097      & 0.181      & 0.312      & 0.528 &  & 0.591     & 0.083      & 0.155      & 0.290      & 0.518 \\
		Monomer     & 0.702     & \textbf{0.169}  & 0.286      & 0.458      & 0.691 &  & 0.705     & 0.176      & 0.289      & 0.457      & 0.690 \\
		BPR-DAE      & 0.709     & 0.167      & 0.273      & 0.467      & 0.704 & & 0.695     & 0.173      & 0.282      & 0.439      & 0.675 \\
		Triplet Net   & 0.706     & 0.163      & 0.280      & 0.457      & 0.696 &  & 0.701     & \textbf{0.181}      & 0.287      & 0.449      & 0.683 \\
		TA-CSN      & 0.716     & 0.167      & 0.284      & 0.467      & 0.708  & & 0.702     & 0.173      & 0.284      & 0.451      & 0.684 \\
		PAICM      & 0.703     & 0.168      & 0.271      & 0.463      & 0.697  & & 0.703     & 0.170      & 0.266      & 0.456      & 0.687 \\
		\cline{1-12}
		\textbf{AFRec}   & \textbf{0.741} & 0.164 & \textbf{0.305} & \textbf{0.500} & \textbf{0.789} & & \textbf{0.753} & 0.180 & \textbf{0.397} & \textbf{0.516} & \textbf{0.828} \\
		\cline{1-12}

	\end{tabularx}
	\label{tab:performances}
\end{table}
 
\subsection{Evaluation Protocols}
For each positive top-bottom pair $(t_i, b_j)$ in the test set, we generate negative test instances by replacing the bottom item $b_j$ with 100 uniformly sampled negative bottom items that are not matched by the top item $t_i$. Then, we choose two commonly-used evaluation metrics, namely HR$@$K and Area Under the ROC Curve (AUC) to compare our model's performance against other baseline methods. HR$@$K indicates the percentage of correctly identified clothes pairs ranked in all top-$K$ lists, which is formulated as the following:
\begin{equation}
	\text {HR} @ K=\frac{\# \text {hit} @ K}{|D_{test}|},
\end{equation}
where $D_{test}$ denotes the collection of all test cases. Meanwhile, AUC is defined as follows:
\begin{equation}
	\text{AUC} = \frac{\sum pred_{positive} > pred_{negative}}{N_{positive} \times N_{negative}},
\end{equation}
where $\sum pred_{positive} > pred_{negative}$ represents the number of test cases that predicted score of positive pairs are larger than negative pairs, while $N_{positive}$ and $N_{negative}$ respectively denote the total number of positive and negative pairs.

\subsection{Implementation Details}
AFRec is implemented using PyTorch \cite{PaszkeGMLBCKLGA19} with Nvidia GTX 2080 Ti. For consistency, we apply the same dimension size $D$ for all embeddings and hidden states. Specifically, we set $D$ to 512. All the trainable parameters in our model are optimised using Adam optimiser \cite{KingmaB14} with the batch size of 64, the learning rate of 1e-4 and the weight decay of 1e-5. To help AFRec converge faster, we first pretrain the SAE module in AfRec using our attribute and category prediction objectives. The attribute classification accuracy of the pretrained SAE module is illustrated in Table \ref{tab:attribute_labels}. As can be seen, the model is highly confident in comprehensively extracting attribute information from visual features, and this allows AFRec to generate meaningful attribute-specific representations for the final recommendation task. After SAE is fully pretrained, we train the whole model in an end-to-end manner.

\subsection{Analysis on Recommendation Effectiveness}
We summarise the evaluation results of all models on the complementary clothing recommendation task with Table \ref{tab:performances}. From the results in the table, we can observe that our AFRec outperforms other state-of-the-art methods on most evaluation metrics, reflecting the effectiveness of our model. This is mainly because our model significantly benefits from the semantic attributes when modelling the compatibility at a fine-grained level. This helps AFRec better capture the complicated interactions among attributes. As a category-unaware model, SiameseNet merely learns fashion compatibility within a unified latent space, and it underperforms due to the lack of the ability to leverage the subtle yet informative attribute signals. 
By incorporating category-awareness in different learning schemes, we can observe similar performance from Monomer, BPR-DAE, Triplet Net and TA-CSN. This indicates that categorical information is helpful for advancing the performance in the task of complementary clothing recommendation. Among these methods, TA-CSN that uses type-specific mask embeddings yields better recommendation accuracy. 
This implies that instead of simply concatenating category embeddings to the global visual embeddings, performing mask operations can let the model focus on certain dimensions of item embeddings for downstream tasks. The attribute-aware method PAICM achieves similar performance to the category-aware methods, which demonstrates the effectiveness of incorporating attribute information for compatibility modelling. However, PAICM models compatibility with a single merged attribute-level embedding for each item. This modelling scheme may fail to capture sufficient disentangled attribute information since all attribute-specific information is fused. In contrast, our model not only accounts for the categorical information via categorical projection spaces, but also mines fine-grained compatibility relations by modelling meaningful semantic attribute interactions.

\begin{table}[t]
	\caption{Performance comparison between different variants of AFRec}
	\setlength\tabcolsep{3.5pt}
	\renewcommand{\arraystretch}{2}
	\begin{tabularx}{\textwidth}{lllllp{1pt}lllr}
		\cline{1-10}
		\multirow{2}{*}{Variants} & \multicolumn{4}{l}{FashionVC} &  & \multicolumn{4}{l}{PolyvoreMaryland}\\
		\cline{2-5} \cline{7-10}
		 & \multirow{2}{*}{AUC} & \multicolumn{3}{l}{HR$@$K} & & \multirow{2}{*}{AUC} & \multicolumn{3}{l}{HR$@$K}\\
		\cline{3-5} \cline{8-10} &
		& K=10 & K=20 & K=40 & & & K=10 & K=20 & K=40 \\
		\cline{1-10}
		AFRec$_{w/o.attr.loss}$ (\ref{eq:attr_loss}) & 0.703 & 0.234 & 0.468 & 0.664 &
		& 0.732 & 0.312 & 0.484 & 0.734\\
		AFRec$_{w/o.cate.loss}$ (\ref{eq:final_loss}) & 0.717 & 0.281 & 0.461 & 0.688 &
		& 0.750 & 0.344 & 0.508 & 0.773\\
		AFRec$_{w/o.attention}$  & 0.703 & 0.172 & 0.461 & 0.703 &
		& 0.749 & 0.344 & 0.515 & 0.758\\
		AFRec$_{self.attention}$  & 0.718  & 0.227 & 0.445 & 0.703 &
		& 0.740 & 0.367 & 0.492 & 0.758\\
		AFRec$_{w/o.cate.projection}$ (\ref{eq:compat_calculation})  & 0.714 & 0.141 & 0.461 & 0.672 &
		& 0.636 & 0.203 & 0.344 & 0.602\\
		AFRec$_{attr.avg}$ & 0.731 & 0.258 & 0.453 & 0.688 &
		& 0.724 & 0.305 & 0.477 & 0.727\\
		
		\cline{1-10}
		Full Version   & 0.741 & 0.305 & 0.500 & 0.789 & & 0.753 & 0.397 & 0.516 & 0.828 \\
		\cline{1-10}

	\end{tabularx}
	\label{tab:ablation_study}
\end{table}

\subsection{Ablation Study}
To verify the contribution of each proposed component in our model, we implement multiple variants of AFRec to perform an ablation study. The evaluation results on both datasets are demonstrated in Table \ref{tab:ablation_study}. We introduce and analyse the effect of each variant of AFRec as follows:
\begin{itemize}
    \item \textbf{AFRec$_{w/o.attr.loss}$}. This variant removes the attribute prediction loss, and all the embeddings of extracted attributes are treated as latent vectors containing different latent global visual information. The obvious performance drop on both datasets indicates that the attribute prediction task can effectively augment the expressiveness of the learned representations. 
    \item \textbf{AFRec$_{w/o.cate.loss}$}. When removing the category classification loss, we can observe mild performance drop on both datasets. Intuitively, we use category classification loss to help the SAE module to concentrate on different parts of fashion items in different categories when learning their global visual features, making our reciprocal attention module highly effective.
    \item \textbf{AFRec$_{w/o.attention}$} and \textbf{AFRec$_{self.attention}$}. We study the contribution of reciprocal attention module by either directly removing the whole attention module (i.e., AFRec$_{w/o.attention}$) or replacing it with a self-attention module (i.e., \textbf{AFRec$_{self.attention}$}) that does not support attribute-wise comparison between different items. We can see a similar performance drop on most evaluation metrics for these two variants. Hence, the results justify that our reciprocal attention effectively avoids the biases caused by the low compatibility scores of unimportant attribute features. 
    \item \textbf{AFRec$_{w/o.cate.projection}$}. AFRec receives the worst evaluation results among all variants when its category-specific projection matrices are removed. This is mainly because the item compatibility measurement varies in different categories. The category-specific projection can let AFRec focus on different latent features of the attributes in different categories.
    \item \textbf{AFRec$_{attr.avg}$}. This variant calculates the compatibility score using only a single embedding vector composed by averaging all attribute embeddings for each item. We can find a slight performance drop on both datasets. This is because the averaged attribute embeddings contain a mixture of multiple attribute characteristics, which may hinder AFRec from making precise compatibility measurement since all attribute factors are entangled. 
\end{itemize}

\begin{figure}[t]
 \centering
 \subfloat[][]
 {\includegraphics[width=0.5\textwidth]{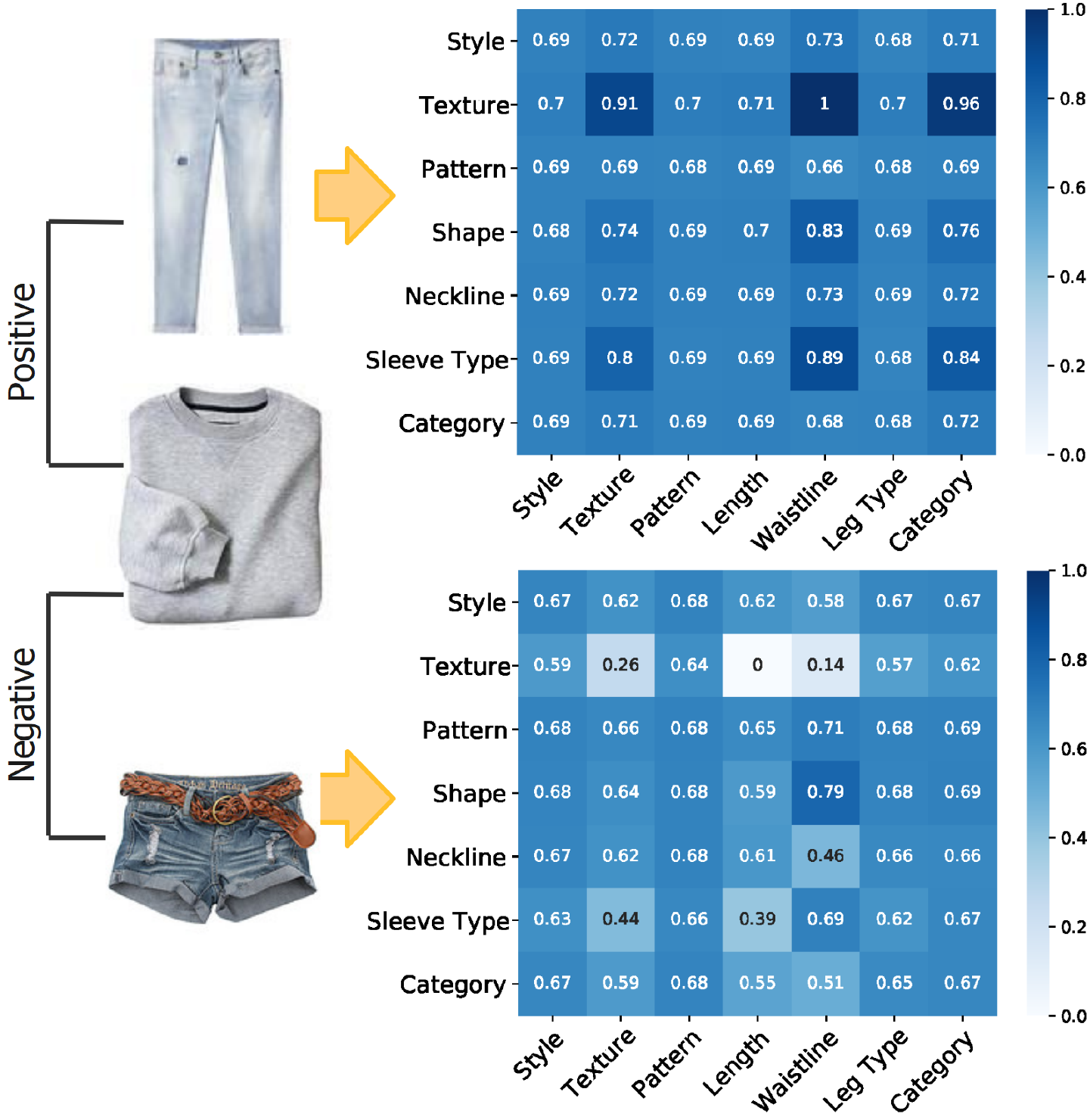}}
 \subfloat[][]
 {\includegraphics[width=0.5\textwidth]{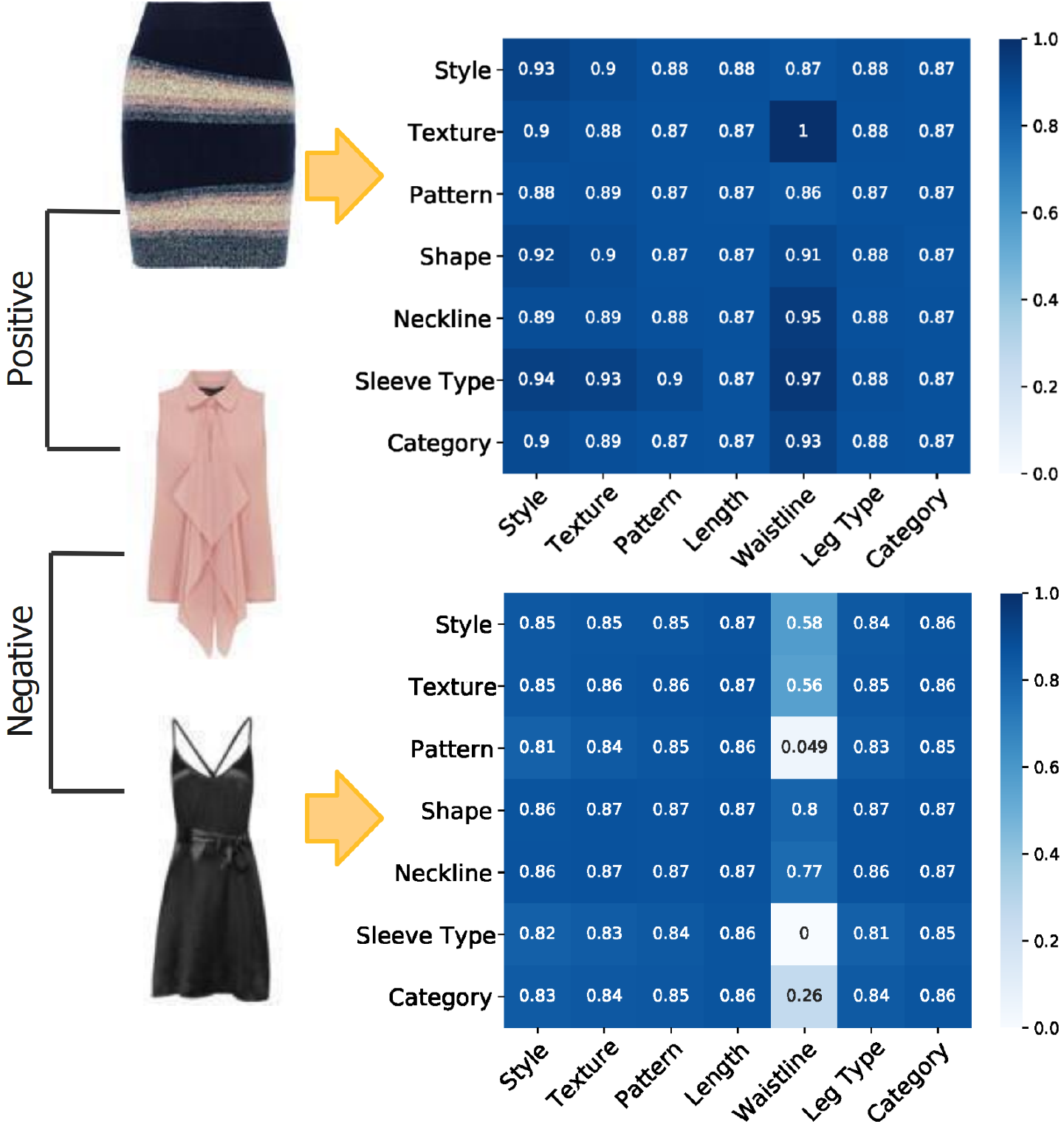}}\\
 \subfloat[][]
 {\includegraphics[width=0.5\textwidth]{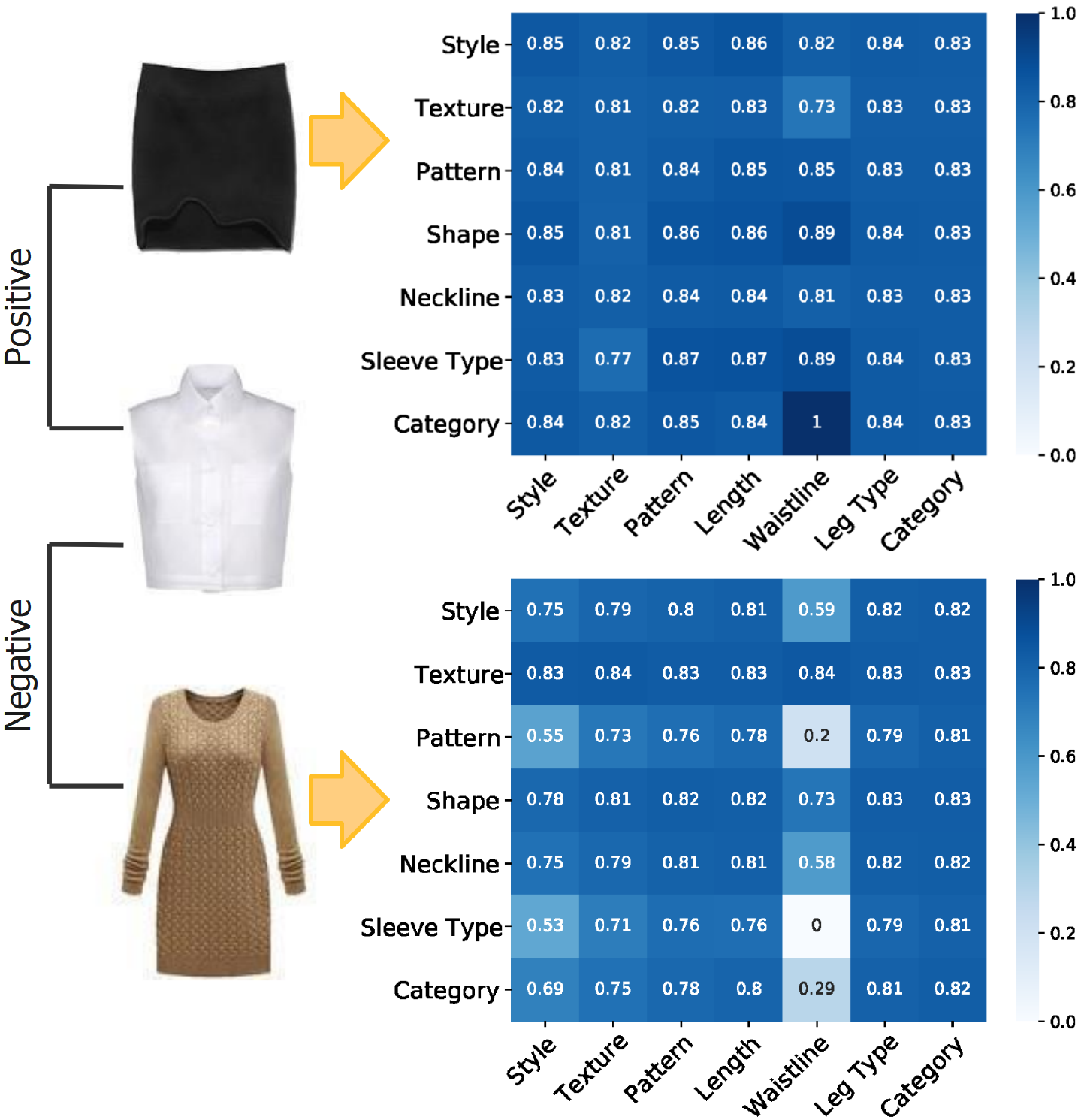}}
 \subfloat[][]
 {\includegraphics[width=0.5\textwidth]{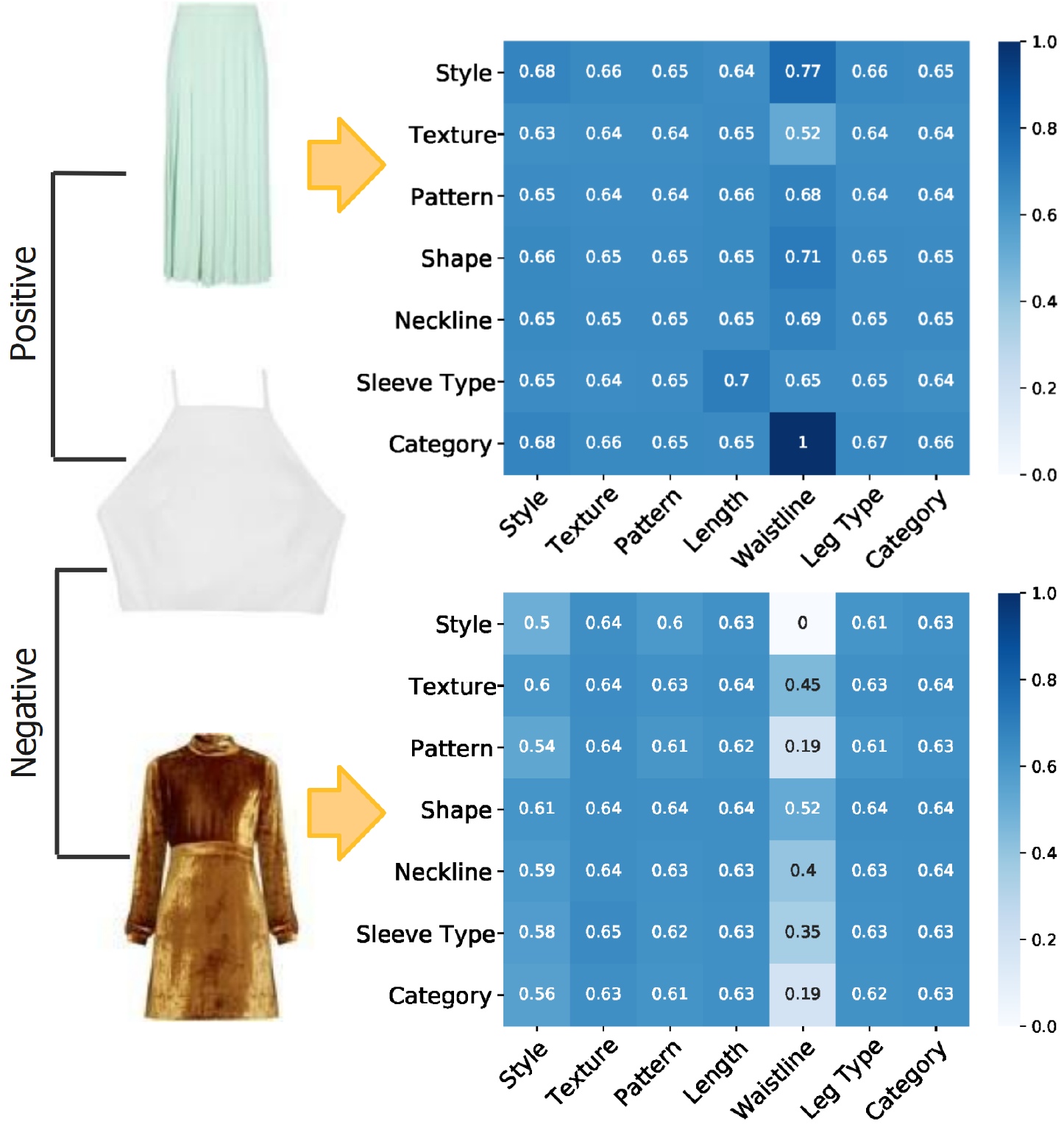}}
 \caption{Visualisation of four pairs of positive and negative outfit test instances}
 \label{fig:heatmap}
\end{figure}

\subsection{Analysis on Recommendation Explainability}
\subsubsection{Visualisation Results}
As attribute-wise compatibility learning plays a crucial role in facilitating the explainability of our model, we select four positive and negative pairs from FashionVC and PolyvoreMaryland dataset, respectively. We use four groups of examples, where each top item is paired with a successfully recommended bottom item and a negative item. We also visualise the computed weighted compatibility matrix $\textbf{M}^{weighted.compat}$ in Figure \ref{fig:heatmap}. Note that each entry in $\textbf{M}^{weighted.compat}$ is rescaled to $[0,1]$ range for better readability.
For instance, in Figure \ref{fig:heatmap}(a), for the positive clothes pair, the values within the compatibility matrix are commonly higher than the negative clothes pair, indicating an overall strong complementary relationship between the grey sweater and the light blue jeans. To name a few, the key matching patterns for two items include the high compatibility between the textures of both items; also, the shape and sleeve type of the sweater is a good fit for the waistline design of the jeans recommended. 

The second observation we can draw from this explainability study is that, for the same top item, its positive match (i.e., a bottom item) commonly performs better in almost all pairwise compatibility between attributes. Also, a positive item tends to exhibit advantages on some key attribute types over the negative item. For example, the compatibility between sleeve type (top) and waistline (bottom) Figure \ref{fig:heatmap}(b) varies significantly in positive and negative pairs. Similar results can be observed from the compatibility between pattern (top) and style (bottom) in Figure \ref{fig:heatmap}(c), and the compatibility between style (top) and waistline (bottom) in Figure \ref{fig:heatmap}(d). To summarise, the attribute-level explanation offers a highly intuitive means for users to understand the reasons why a pair of clothes are complementary or not.
The explainability makes it easier to provide people with insights into which attribute factors are the main contributors in clothing matching.

\subsubsection{User Study}
We further conduct a user study based on 10 randomly selected volunteers (5 are male and 5 are female) to quantitatively evaluate the utility of our generated explanations to real users. Specifically, we first use our model and the explainable baseline method PAICM \cite{HanSYWN19} to generate the prediction and interpretation results for uniformly sampled 100 clothing outfits consisting of 50 positive and 50 negative pairs. In the user study, each participant is provided with all 100 visualisation results, and are asked to up-vote or down-vote the explanations generated by AFRec and PAICM. We collected responses from all participants, and report the up-vote ratio with Table \ref{tab:user_study}. 
On the positive test instances, both methods generate decent explanations on what attribute factors are critical for the harmony of an outfit. In negative instances, the explanations generated by AFRec are more preferred. From the participants' responses, they mainly agree on the incompatible category, style, and texture attributes identified by AFRec. The key reason for better explainability of our model is that modelling interactions among attributes under category-specific spaces can benefit the model to capture more details between two items. In comparison, PAICM merges the attribute information into a single embedding, which may neglect some subtle information contained in images leading to low-quality prototype embeddings. As a result, those prototype embeddings may mislead the model to provide wrong explanations.

\begin{table}[t]
\caption{Up-vote ratio of the generated explanations}
\renewcommand{\arraystretch}{2}
\begin{tabularx}{\textwidth}{p{0.25\textwidth}p{0.25\textwidth}p{0.25\textwidth}p{0.25\textwidth}}
\hline
                             & Model & Positive & Negative \\ \hline
\multirow{2}{*}{Up-vote Ratio ($\%$)} & AFRec & \textbf{66.0}     & \textbf{48.0} \\
                             & PAICM &  64.0    &   38.0    \\ \hline
\end{tabularx}
\label{tab:user_study}
\end{table}


\section{Conclusion}
To deal with the lack of explainability of existing complementary clothing recommendation approaches, we propose a novel solution named AFRec in this paper. AFRec obtains attribute-specific representations from fashion items by a CNN-based attribute embedding extractor to support fine-grained fashion compatibility modelling and enhances its explainability towards the prediction results. Our experiments on two large-scale benchmark datasets show the effectiveness and interpretability of AFRec, demonstrating the strong practicality in real-life scenarios.
%
%
\acknowledgement{This work is partially supported by Australian Research Council Discovery Project (ARC DP190102353, DP190101985, CE200100025).}

\bibliographystyle{spmpsci}      
\bibliography{wwwj21}   

\end{document}